\documentclass{elsart}

\usepackage{graphicx}

\begin{document}

\begin{frontmatter}

\title{On the Influence of Uniform Singularity Distributions}

\author{S.Mukhopadhyay, N.Majumdar}
\address{Saha Institute of Nuclear Physics\\
1/AF, Sector 1, Bidhannagar, Kolkata 700064\\
West Bengal, India\\
e-mail: supratik.mukhopadhyay@saha.ac.in
}

\begin{abstract}
Exact expressions for three-dimensional potential and force field due to
uniform singularity distributed on a finite flat rectangular surface have been
presented. The expressions, valid throughout the physical domain, have been
found to be consistent with other expressions available as special cases of the
same problem. Very accurate estimates of the capacitance of a unit square plate
and a unit cube have been made using them.
\end{abstract}


\end{frontmatter}

\newpage

\section{Introduction}
The importance of inverse laws has been acknowledged in various branches of
science and technology for a long time. Much of the contemporary physics is
also dominated by the effects of inverse laws in various guises. Whenever, a
particular physical phenomenon is modeled using sources or sinks, the inverse
laws come into play. These laws are found to be crucially important in
gravitation, electromagnetics, ideal fluid dynamics, Stoke's flow, acoustics,
optics,
thermodynamics and many other fields. In fact, a large part of the classical
physics, when assumed non-dissipative, can be described by some form of the
inverse laws such as the Laplace's and Poisson's equations. These two linear
second order partial differential equations have been considered to be among
the most important differential equations in the whole of classical physics.
As a result, estimation of the effects of the inverse laws has been known to be
extremely important in many branches of science and technology \cite{Feynman}.

While the effect of point sources and sinks can be easily computed, it has not
been possible to obtain closed form expressions for computing the effects of
distributed sources, except for very simple cases. But, since in many
of the real-life problems the singularities are found to
be distributed on surfaces of various shapes and sizes, it has been customary to
represent them using the simple shapes for which closed form expressions are
known, or simply by assuming the surface to be composed of a large number of
point sources. These approximations, besides being computationally rather
expensive, turn out to be significantly restricted and inaccurate.

In this work, we have presented the closed form expressions of potential
and field due to a uniform distribution of source on a flat surface. Using these
expressions it has been possible to find the potential and field accurately in
the complete physical domain, including the critical near-field domain.
Especially, the
sharp changes and discontinuities which characterize the near-field domain have
been easily reproduced. As a result of the numerical experiments carried out to
establish the accuracy of the proposed method, it has been possible to estimate
the amount of discretization required to predict the field properties up to a
desired level of accuracy.

Although in the present work we have concentrated only on source distribution on
flat surfaces, surfaces of other shapes, including curved ones, can be similarly
handled through the use of proper geometric transformations. Similar expressions
may also be used in dynamic situations where the assumption of quasi-static
holds. Since the expressions are analytic and valid for the complete physical
domain, and no approximations regarding the size or shape of the singular
surface have been made during their derivation, their application is not
limited by the proximity of other singular surfaces or their curvature.
In fact, it has been well-known that most of the difficulties in the
earlier methods arose because of nodal concentration of singularities which led
to various mathematical difficulties and to the infamous numerical boundary
layers \cite{Sladek91,Chyuan04}. Through the use of the expressions presented
in this work, it is possible to truly model the effect of distributed
singularities precisely. As a result, the problem of mathematical singularities
does not arise at all, and the real, physical singularities can be dealt with
in a straight-forward manner. Moreover, the requirement of developing special
formulations (despite their mathematical elegance and efficiency) in order to
cope with the singularities in various guises does not arise.

As an illustration of the use of the presented expressions, we have developed
a boundary element method solver, namely, the Nearly Exact Boundary Element
Method (NEBEM) solver. Using this solver, we have computed the capacitances of a
unit square plate and a unit cube to very high accuracy. These problems have
been considered to be two major unsolved problems of electrostatic theory
\cite{Maxwell,Reitan,Solomon,Goto,Douglas,Read,Given,Mansfield,Hwang03,Hwang04,Mascagni,Wintle}
and no analytical solutions for these problems have been obtained so far.
The capacitance values estimated by the present method have been compared with
very accurate results available in the literature (using boundary element
methods (BEM) and others).
The comparison testifies to the accuracy of the NEBEM solver and, hence, the
usefulness of the presented expressions.

\section{Theory}
\begin{figure}[hbt]
\begin{center}
\includegraphics[height=2in,width=3in]{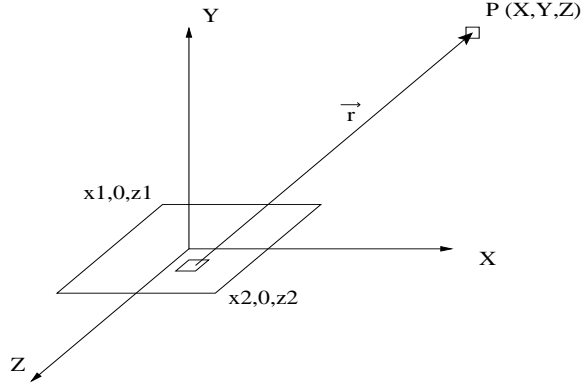}
\caption{\label{fig:GeomElem} A rectangular surface with uniform distributed
source}
\end{center}
\end{figure}

The expression for potential at a point $(X, Y, Z)$ in free space due to uniform
source distributed on a rectangular flat surface having corners situated at
$(x_1, z_1)$ and $(x_2, z_2)$ as shown in Fig.\ref{fig:GeomElem} is known to
be a multiple of
\begin{equation}
\phi(X,Y,Z) = \int_{z_1}^{z_2} \int_{x_1}^{x_2}
			\frac{dx\,dz}{\sqrt{(X-x)^2 + Y^2 + (Z-z)^2}}
\label{eqn:PotInt}
\end{equation}
where the value of the multiple depends upon the strength of the source and
other physical considerations. Here, the surface under consideration, as well as
the origin of the coordinate system is on the $XZ$ plane.
The denominator within the integrals can be
easily interpreted to be the distance between the point $(X,Y,Z)$ at which the
potential is being evaluated and an infinitesimal element on the surface.
The closed form expression for $\phi(X,Y,Z)$ is as follows:
\begin{eqnarray}
\label{eqn:PotExact}
\lefteqn{\phi(X,Y,Z) =} \nonumber \\
&& (X-x_1)\,\ln \left( \frac{D_{12}\, -\, (Z-z_2)} {D_{11}\, -\, (Z-z_1)} \right) + (X-x_2)\,\ln \left( \frac{D_{21}\, -\, (Z-z_1)} {D_{22}\, -\, (Z-z_2)} \right) \nonumber \\
&& + (Z-z_1)\,\ln \left( \frac{D_{21}\, -\, (X-x_2)} {D_{11}\, -\, (X-x_1)} \right) + (Z-z_2)\,\ln \left( \frac{D_{12}\, -\, (X-x_1)} {D_{22}\, -\, (X-x_2)} \right) \nonumber \\
&& + \frac{i\, |Y|}{2} \nonumber \\
&& \left( \right. S_1\, \left( \right.
tanh^{-1} \left( \frac {R_1 + i\, I_1} {D_{11}\, \left| Z - z_1 \right|}  \right)
-\, tanh^{-1} \left( \frac {R_1 - i\, I_1} {D_{11}\, \left| Z - z_1 \right|}  \right) \nonumber \\
&& +\, tanh^{-1} \left( \frac {R_1 - i\, I_2} {D_{21}\, \left| Z - z_1 \right|}  \right)
+\, tanh^{-1} \left( \frac {R_1 + i\, I_2} {D_{21}\, \left| Z - z_1 \right|}  \right) \left. \right) \nonumber \\
&& +\, S_2\, \left( \right.
tanh^{-1} \left( \frac {R_2 + i\, I_2} {D_{22}\, \left| Z - z_2 \right|}  \right)\,
-\, tanh^{-1} \left( \frac {R_2 - i\, I_2} {D_{22}\, \left| Z - z_2 \right|}  \right) \nonumber \\
&& +\, tanh^{-1} \left( \frac {R_2 + i\, I_1} {D_{21}\, \left| Z - z_2 \right|}  \right)\,
-\, tanh^{-1} \left( \frac {R_2 - i\, I_1} {D_{21}\, \left| Z - z_2 \right|}  \right) \left. \right) \left. \right) \nonumber \\
&& - 2\,\pi\,Y
\end{eqnarray}
where
\begin{eqnarray*}
D_{11} = \sqrt { (X-x_1)^2 + Y^2 + (Z-z_1)^2 };
D_{12} = \sqrt { (X-x_1)^2 + Y^2 + (Z-z_2)^2 } \\
D_{21} = \sqrt { (X-x_2)^2 + Y^2 + (Z-z_1)^2 };
D_{22} = \sqrt { (X-x_2)^2 + Y^2 + (Z-z_2)^2 } \\
R_1 = Y^2 + (Z-z_1)^2;
R_2 = Y^2 + (Z-z_2)^2 \\
I_1 = (X-x_1)\,\left| Y \right|;
I_2 = (X-x_2)\,\left| Y \right|;
S_1 = {\it sign} (z_1-Z);
S_2 =  {\it sign} (z_2-Z)
\end{eqnarray*}

Similarly, the force field for the above problem is given as a multiple of
\begin{equation}
\label{eqn:FInt}
\vec{F}(X,Y,Z) = \int_{z_1}^{z_2} \int_{x_1}^{x_2}
            \frac{\hat{r}\,dx\,dz}{r^2}
\end{equation}
where $\vec{r}$ is the displacement vector from a small surface element to the
$(X,Y,Z)$ point where the force field is being evaluated.
Eqn.(\ref{eqn:FInt}) has also been integrated in order to get exact expressions
to estimate the force fields in the X, Y and Z directions. These expressions,
valid for the complete physical domain, are as follows:
\begin{equation}
\label{eqn:FxExact}
F_x(X,Y,Z) =
\ln \left( \frac{D_{11}\, -\, (Z-z_1)} {D_{12}\, -\, (Z-z_2)} \right) \,+\, \ln \left( \frac{D_{22}\, -\, (Z-z_2)} {D_{21}\, -\, (Z-z_1)} \right)
\end{equation}
\begin{eqnarray}
\label{eqn:FyExact}
\lefteqn{F_y(X,Y,Z) =} \nonumber \\
&& -\, \frac{i}{2}\, Sign(Y) \nonumber \\
&& \left( \right. S_1\, \left( \right.
tanh^{-1} \left( \frac {R_1 + i\, I_1} {D_{11}\, \left| Z - z_1 \right|}  \right)
-\, tanh^{-1} \left( \frac {R_1 - i\, I_1} {D_{11}\, \left| Z - z_1 \right|}  \right) \nonumber \\
&& +\, tanh^{-1} \left( \frac {R_1 - i\, I_2} {D_{21}\, \left| Z - z_1 \right|}  \right)
+\, tanh^{-1} \left( \frac {R_1 + i\, I_2} {D_{21}\, \left| Z - z_1 \right|}  \right) \left. \right) \nonumber \\
&& +\, S_2\, \left( \right.
tanh^{-1} \left( \frac {R_2 + i\, I_2} {D_{22}\, \left| Z - z_2 \right|}  \right)\,
-\, tanh^{-1} \left( \frac {R_2 - i\, I_2} {D_{22}\, \left| Z - z_2 \right|}  \right) \nonumber \\
&& +\, tanh^{-1} \left( \frac {R_2 + i\, I_1} {D_{21}\, \left| Z - z_2 \right|}  \right)\,
-\, tanh^{-1} \left( \frac {R_2 - i\, I_1} {D_{21}\, \left| Z - z_2 \right|}  \right) \left. \right) \left. \right) \nonumber \\
&& +\, \it{C}
\end{eqnarray}
\begin{equation}
\label{eqn:FzExact}
F_z(X,Y,Z) =
\ln \left( \frac{D_{11}\, -\, (X-x_1)} {D_{21}\, -\, (X-x_2)} \right) \, + \,\ln \left( \frac{D_{22}\, -\, (X-x_2)} {D_{12}\, -\, (X-x_1)} \right)
\end{equation}
In Eqn.(\ref{eqn:FyExact}), $C$ is a constant of integration as follows:
\[\it{C} = \left\{
\begin{array}{l l}
  0 & \quad \mbox{if outside the extent of the flat surface}\\
  2\, \pi & \quad \mbox{if inside the extent of the surface and Y $>$ 0}\\
  -2\, \pi & \quad \mbox{if inside the extent of the surface and Y $<$ 0} \end{array} \right. \]

The above closed-form integrations can be useful in the mathematical modeling
of physical processes governed by the inverse square laws as designated by
Eqns.(\ref{eqn:PotInt}) and (\ref{eqn:FInt}).
Eqns. (\ref{eqn:PotExact}) and (\ref{eqn:FxExact})-(\ref{eqn:FzExact}),
being exact and valid throughout the physical domain, can be used to
formulate and solve multi-scale problems involving Dirichlet, Neumann or Robin
boundary conditions.

It is well known that the potential at the centroid of a flat rectangular
surface of same dimensions and having source uniformly distributed on it is as
follows:
\begin{equation}
\label{eqn:PotCentroid}
\phi(0,0,0) = 2 ( a\, \log(\frac{\sqrt{a^2+b^2} + b}{a})
+ b\, \log(\frac{\sqrt{a^2+b^2} + a}{b}) )
\end{equation}
where $a$ and $b$ are the sides of the rectangular surface. Considering the
coordinate origin at the centroid of the element shown in Fig.\ref{fig:GeomElem}
and $x_2 - x_1 = a$, $z_2 - z_1 = b$,
by simple algebraic manipulation we can easily show that the expression for
potential given by Eqn.(\ref{eqn:PotExact}) reduces to
Eqn.(\ref{eqn:PotCentroid}).
In addition, if the point $P(X,Y,Z)$ is infinitely far away from the surface on
which the singularity is distributed, the following relations can be easily
shown to be true:
\begin{eqnarray*}
X = X - x_1 = X - x_2 \\
Z = Z - z_1 = Z - z_2 \\
I_1 = I_2 = X |Y| \\
R_1 = R_2 = Y^2 + Z^2 \\
D_{11} = D_{12} = D_{21} = D_{22} = \sqrt{X^2 + Y^2 + Z^2}
\end{eqnarray*}
Substituting the above in Eqn.(\ref{eqn:PotExact}), we find that at a point
infinitely far away from the influencing element, the potential $\phi$ is zero.
Thus, the presented expressions have been found to be consistent with results
in the near-field, as well as the far-field.

As discussed in the Introduction, the above expressions (with suitable
constants) have been used to compute the capacitances of a unit square plate and
a unit cube in the framework of BEM. The resulting BEM solver has been
christened as the Nearly Exact BEM (NEBEM) solver. The details of the solver
have been presented in several recent communications, e.g.,
\cite{Mukhopadhyay05} and we will refrain from elaborating on it here.

\section{Results}
In order to establish the accuracy of the expressions in
Eqns.(\ref{eqn:PotExact}) and (\ref{eqn:FxExact})-(\ref{eqn:FzExact}), we have
computed the potential and field distributions of a unit flat square
($1unit \times 1unit$) conducting plate carrying uniform unit singularity density
($1\, unit/m^2$). Results computed using the exact expressions
have been compared with those computed using a conventional zero-th order
piecewise constant BEM with varying amount of discretization ($1 \times 1$,
$10 \times 10$, $100 \times 100$ and $1000 \times 1000$ elements). For the
line plots below, we have presented computations either along a diagonal
which originates at $(-1.5, -1.5, -1.5)$ and ends at $(1.5, 1.5, 1.5)$ (please
refer to Fig.\ref{fig:GeomElem}), or one which runs parallel to X- or Z-axis
very close to the edge of the element, just 10 nanounits away from it.
\begin{itemize}
\item
{Potential along the diagonals and edges: 
From the Figs.\ref{fig:PotDiag} and \ref{fig:PotEdge}, it is clear that the
usual BEM solver produces acceptable
results only after the plate is discretized into $100 \times 100$ elements.
The observation is true along the diagonal line, as well as the line along
the edge. For the latter, an oscillation in the usual BEM results is observed
for all the discretizations while the exact expressions produce quite smooth
results.}
\begin{figure}[hbt]
\begin{center}
\includegraphics[height=2in,width=3in]{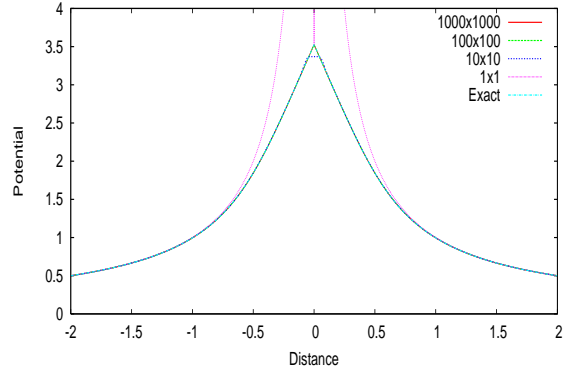}
\caption{\label{fig:PotDiag} Comparison of potential distribution along diagonal.}
\end{center}
\end{figure}
\begin{figure}
\begin{center}
\includegraphics[height=2in,width=3in]{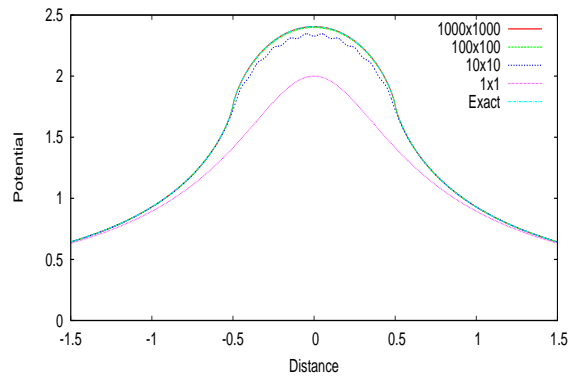}
\caption{\label{fig:PotEdge} Comparison of potential distribution along edge.}
\end{center}
\end{figure}

\item
{Force components along the diagonals and edges:
The $X$ and $Y$ force components for the above two cases have been presented
in Figs.\ref{fig:FxDiag}, \ref{fig:FyDiag} and \ref{fig:FxEdge}.
Comments made above are equally applicable in these cases,
only in a more significant manner. This is expected because forces are
obtained as gradients of potential and, as a result, the short-comings in the
computation of potential are amplified in the estimation of force components.
As a result, even a $1000 \times 1000$ discretization fails to yield
satisfactory results. The results from the exact expressions are excellent
throughout.}
\begin{figure}[hbt]
\begin{center}
\includegraphics[height=2in,width=3in]{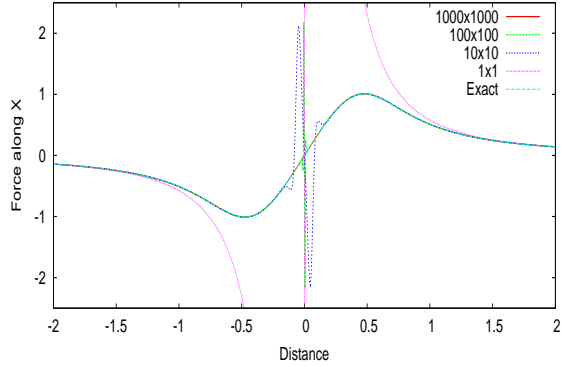}
\caption{\label{fig:FxDiag} Comparison of $F_x$ distribution along diagonal.}
\end{center}
\end{figure}
\begin{figure}[hbt]
\begin{center}
\includegraphics[height=2in,width=3in]{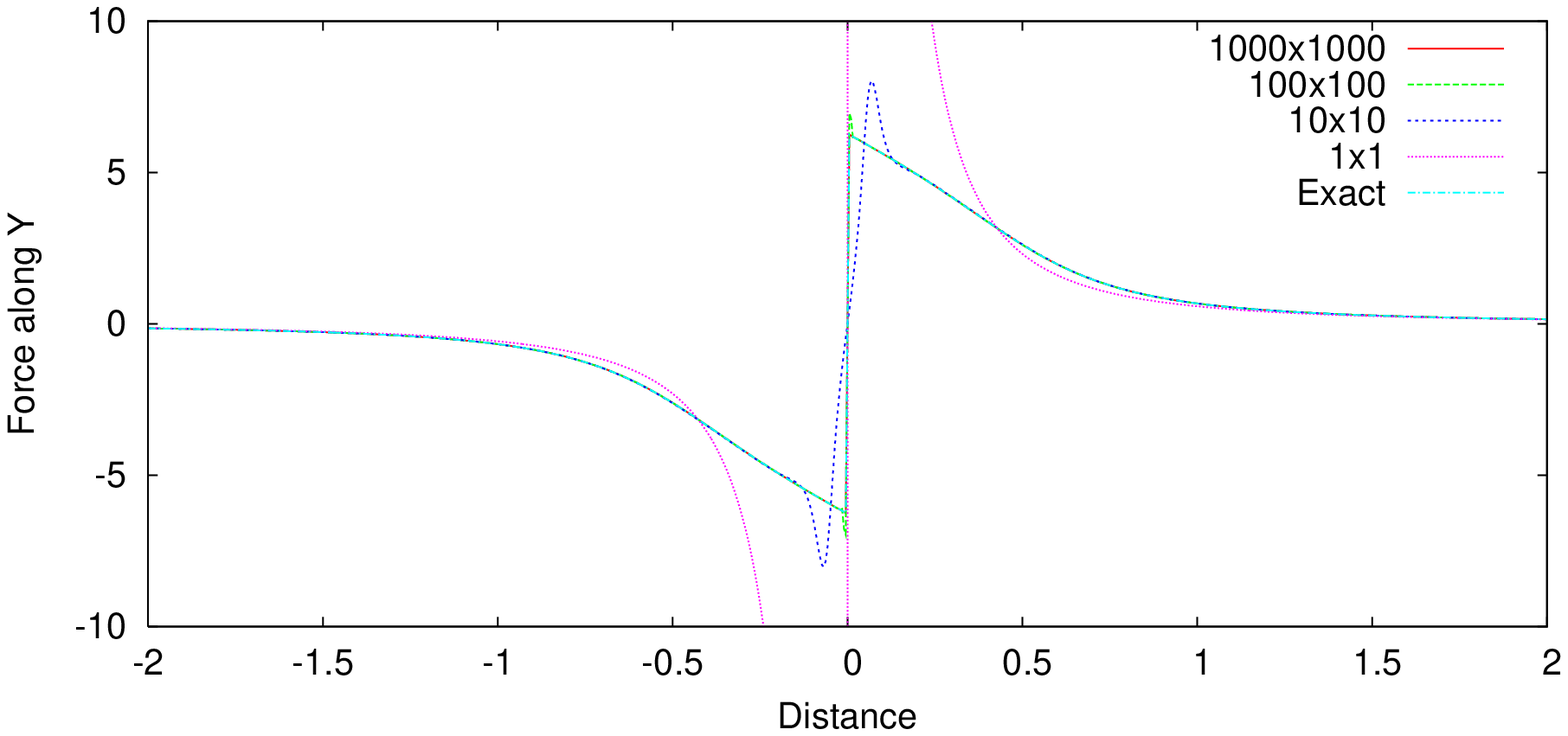}
\caption{\label{fig:FyDiag} Comparison of $F_y$ distribution along diagonal.}
\end{center}
\end{figure}
\begin{figure}[hbt]
\begin{center}
\includegraphics[height=2in,width=3in]{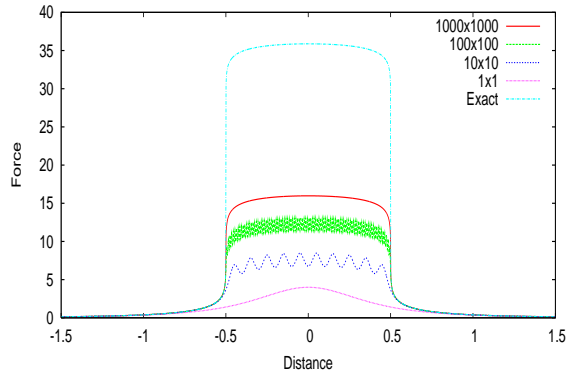}
\caption{\label{fig:FxEdge} Comparison of $F_x$ along the edge.}
\end{center}
\end{figure}

\item
{Surface plot of potential:
To help visualization of the potential field, we have presented a surface
plot of the potential on the conducting plate in Fig.\ref{fig:PotSurf}. Please
note that these are values on the plate itself.}
\begin{figure}[hbt]
\begin{center}
\includegraphics[height=2in,width=3in]{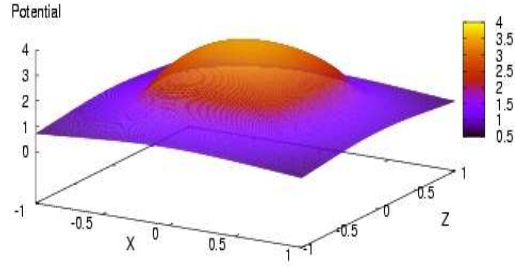}
\caption{\label{fig:PotSurf} Potential distribution on the plate.}
\end{center}
\end{figure}

\item
{Surface plot of force components:
Force surfaces in the $X$ and $Y$ directions are presented in
the following Figs.\ref{fig:FxSurf} and \ref{fig:FySurf}. These force components
have been computed at a distance of only $10nanounits$ from the surface of the
square
plate. The sharp changes in the magnitude of these force components are found
to be accurately estimated by the new expressions.}
\begin{figure}[hbt]
\begin{center}
\includegraphics[height=2in,width=3in]{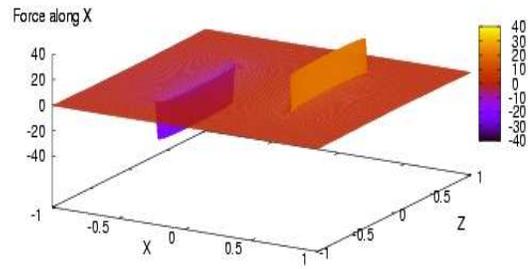}
\caption{\label{fig:FxSurf} $F_x$ distribution at $Y=10nano units$}
\end{center}
\end{figure}
\begin{figure}[hbt]
\begin{center}
\includegraphics[height=2in,width=3in]{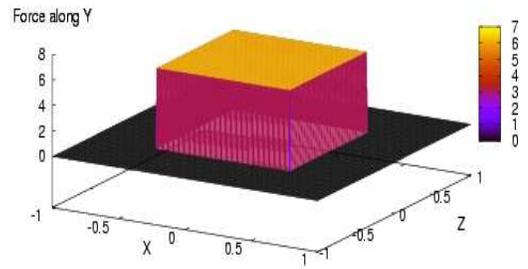}
\caption{\label{fig:FySurf} $F_y$ distribution at $Y=10nano units$}
\end{center}
\end{figure}

\item
{Error plot along diagonals for potential and force components:
In Figs. \ref{fig:ErrorPot} and \ref{fig:ErrorFy}, we have presented the
normalized error, defined as $(Approximate\,-\,Exact)/Exact$ as a function of
distance. One important fact that is immediately apparent is that the
error in force field computation is larger by almost an order of magnitude than
that in the computation of potential. Moreover, it is clear that the usual BEM
approximation ($1 \times 1$) is unacceptable as soon as the distance of the
evaluation point is of the same order of magnitude as the size of the element.
For potential calculations, the error exceeds $1\%$, while for force field
it exceeds $10\%$ even when the distance is twice the size of the plate. At
lesser distances, the error increases rapidly. Only after the unit plate is
further segmented into $100\times100$ elements, the computed results are
acceptable (error less than $1\%$) throughout the domain. But, under the
framework of usual BEM, this translates in to $10000$ elements in place of $1$!
Use of Gaussian quadrature can alleviate the problem to a certain extent, but
is unlikely to provide a complete cure.}

\begin{figure}[hbt]
\begin{center}
\includegraphics[height=2in,width=3in]{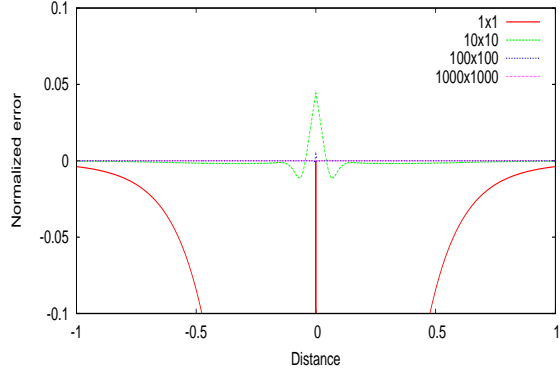}
\caption{\label{fig:ErrorPot} Variation of Error in computing potential along a diagonal.}
\end{center}
\end{figure}
\begin{figure}[hbt]
\begin{center}
\includegraphics[height=2in,width=3in]{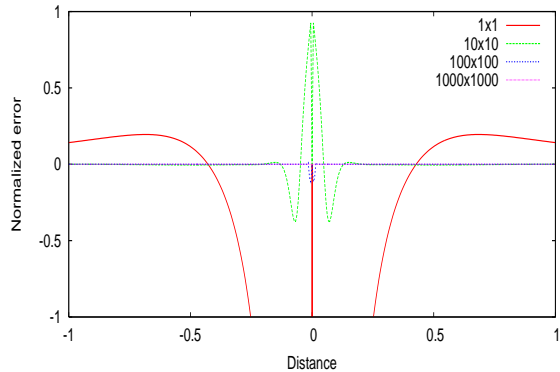}
\caption{\label{fig:ErrorFy} Variation of Error in computing force in Y along a diagonal.}
\end{center}
\end{figure}

\end{itemize}

The above results prove the precision of the closed-form integrals given in
Eqn.(\ref{eqn:PotExact}) and Eqns.(\ref{eqn:FxExact})-(\ref{eqn:FzExact}) which
the NEBEM solver uses as its foundation expressions. As a further application
of them, we have computed the capacitance of a unit square plate and a unit
square cube using the NEBEM solver. In Table\ref{table:CapComp}, we have
presented a comparison of the values of capacitances as calculated by
\cite{Maxwell,Reitan,Solomon,Goto,Douglas,Read,Given,Mansfield,Hwang03,Hwang04,Mascagni}
and our estimations. In \cite{Douglas,Given,Mansfield}, the authors calculated
the capacitances of the plate and cube (among many other things) using numerical
path integration method, while in \cite{Hwang03,Hwang04,Mascagni}, the authors
used the "random walk on the boundary" method. In
\cite{Maxwell,Reitan,Solomon,Goto,Read}, various forms of the BEM were used to
solve the problem. In \cite{Goto,Read}, the authors improved upon the BEM quite
substantially by proposing new and more accurate expressions for the evaluation
of the elements in the influence matrix. However, these expressions
are not valid for the entire physical domain and thus, necessitated the use of
several expressions for evaluating potential and field on the charged surfaces
and other field points. It was also necessary to subdivide the segments on the
boundary to satisfy certain approximations used to deduce the expressions
\cite{Renau}.
Despite these limitations, it must be said that the improvement was significant
through the use of these expressions and their results belong to the most
accurate ones available for the present problem. In order to enhance
the accuracy of the results \cite{Goto,Read,Hwang04} also used extrapolation
techniques. Finally, in \cite{Wintle}, new upper and lower bounds of the
capacitance of an isolated cube have been presented using random walk methods.
New estimates for the two capacitances (preliminary for the plate) have also
been presented here.
In addition, some remarks on the usual BEM have been made which, although
true in general, does not apply to the NEBEM solver. It may be noted here that
our results, as presented in Table\ref{table:CapComp} have been obtained
through a straight-forward application of the NEBEM solver, without taking
recourse to any extrapolation technique. In order to model the sharp change in
the charge density near the corners and edges, finer discretization in these
regions have been used. Although it is difficult to comment regarding which is
the best
result among the published ones, it is clear from the table that the new
expressions indeed lead to very accurate results which are well within the
acceptable range.

\begin{table}[hbt]
\centering
\caption{\label{table:CapComp}Comparison of capacitance values}
\begin{tabular}{| l | l | c | c |}
\hline
Reference & Method & Plate  & Cube \\
\hline
\cite{Maxwell} & Surface Charge & 0.3607 & - \\
\hline
\cite{Reitan} & Surface Charge & 0.362 & 0.6555 \\
\hline
\cite{Solomon} & Surface Charge & 0.367 & - \\
\hline
\cite{Goto} & Refined Surface Charge & $0.3667892 \pm 1.1 \times 10^{-6}$ & $0.6606747 \pm 5 \times 10^{-7}$\\
& and Extrapolation & & \\
\hline
\cite{Douglas} & Brownian Dynamics & - & 0.663 \\
\hline
\cite{Read} & Refined Boundary Element & $0.3667874 \pm 1 \times 10^{-7}$ & $0.6606785 \pm 6 \times 10^{-7}$\\
& and Extrapolation & & \\
\hline
\cite{Given} & Refined Brownian Dynamics & - & $0.660675 \pm 1 \times 10^{-5}$ \\
\hline
\cite{Mansfield} & Numerical Path Integration & 0.36684 & 0.66069 \\
\hline
\cite{Hwang03} & Walk on Spheres & - & $0.660683 \pm 5 \times 10^{-6}$\\
\hline
\cite{Hwang04} & Modified Walk on Spheres & - & $0.6606867 \pm 1.2 \times 10^{-6}$ \\
\hline
\cite{Mascagni} & Random Walk on the Boundary & - & $0.6606780 \pm 2.7 \times 10^{-7}$ \\
\hline
\cite{Wintle} & Random Walk & $0.36 \pm 0.01$ & $0.6606 \pm 0.0001$ \\
\hline
This work & NEBEM & 0.3667869 & 0.6606746 \\
\hline
\end{tabular}
\end{table}

Besides obtaining extremely accurate results,
there are several other advantages of using the presented expressions for
computation of potential and field, in general. For example, for the NEBEM
solver, the boundary condition on a
particular element need not be located only at the centroid of the element. It
can be anywhere on the element, except probably on the edges, or the point
can also be situated very close to the singular surface, while not being exactly
on it. These capabilities can be of significant advantages in specific
situations. In addition, since the singularities are now truly distributed on
the surface, and there is no concept of nodal point in the formulation, the
aspect ratio (ratio of length to breadth) of the elements can vary as wildly
as necessary. The question of mathematical singularities, and the resulting
numerical boundary layer, does not arise at all. As a result, it should be
straight-forward to deal with real physical singularities such as close
proximity of two singular surfaces, or those due to degeneracy of singular
surfaces. All these advantages, along with the fact that the inverse square is
an almost ubiquitous feature of the natural world, makes the expressions
presented in this work particularly suitable for multi-physics multi-scale
problems.

\section{Conclusions}
Exact expressions for potential and force field due to uniform singularity
distribution on a flat surface has been presented. The expressions have been
found to yield very accurate results in the complete physical domain. Of special
importance is their ability to reproduce the complicated field structure in the
near-field region. The errors inducted in assuming discrete point sources to
represent a continuous distribution have been illustrated. Accurate estimates
of the capacitance of a unit square plate and that of a unit cube have been made
using the Nearly Exact BEM (NEBEM) solver which uses Eqns.(\ref{eqn:PotExact}),
(\ref{eqn:FxExact})-(\ref{eqn:FzExact}) as its foundation
expressions. Comparison of the obtained results with very accurate results
available in the literature has confirmed the accuracy of the closed-form
integrals.

\section{Acknowledgments}
We would like to thank Professor Bikash Sinha, Director, SINP and Professor
Sudeb Bhattacharya, Head, NAP Division, SINP for their encouragement and
support throughout the work.

\end{document}